# Cross-probing van der Waals Multiferroics

Bo Peng[1,2,9]*, Zeya Li[3,4,9], Han Wang[1,2], Shengdong You[1,2], Yangliu Wu[1,2], Wei Liu[1,2], Jianliang Xie[1,2], Haipeng Lu[1,2], Peiheng Zhou[1,2], Kian Ping Loh[5]*, Je-Geun Park[6, 7]*, Hongtao Yuan[3,8]* and Longjiang Deng[1,2]*

[1]*National Engineering Research Center of Electromagnetic Radiation Control Materials, School of Electronic Science and Engineering, University of Electronic Science and Technology of China, Chengdu, China.*

[2]*Key Laboratory of Multi Spectral Absorbing Materials and Structures of Ministry of Education, School of Electronic Science and Engineering, University of Electronic Science and Technology of China, Chengdu, China.*

[3]*National Laboratory of Solid State Microstructures, College of Engineering and Applied Sciences, and Collaborative Innovation Center of Advanced Microstructures, Nanjing University, Nanjing 210000, China.*

[4]*Quantum Science Center of Guangdong-Hong Kong-Macao Greater Bay Area, Shenzhen 518045, China.*

[5]*Department of Chemistry, National University of Singapore, Singapore 117543, Singapore.*

[6]*Department of Physics and Astronomy & Institute of Applied Physics, Seoul National University, Seoul 08826, South Korea*

[7]*Center for Quantum Materials, Seoul National University, Seoul 08826, South Korea*

[8]*Jiangsu Physical Science Research Center, Nanjing 210093, China*

[9]*These authors contributed equally: Bo Peng, and Zeya Li*

*Corresponding to: bo_peng@uestc.edu.cn, chmlohkp@nus.edu.sg, jgpark10@snu.ac.kr, htyuan@nju.edu.cn, denglj@uestc.edu.cn*

**Abstract: Multiferroic materials that exhibit coupled electric and magnetic orders have attracted sustained interest since the early proposals of magnetoelectric coupling, driven by their potential to enable low-power memory and logic devices with unprecedented performance. The recent emergence of two-dimensional (2D) van der Waals (vdW) multiferroics has opened exciting new frontiers in this already burgeoning field, offering a platform to explore magnetism, ferroelectricity, and their cross-coupling in atomically thin systems. A particular notable example is a few-layer $NiI_2$ multiferroic, which exhibits coexisting chiral spin textures and (anti)ferroelectricity, accompanied by strong magnetoelectric coupling. Such systems simultaneously break time-reversal and spatial-inversion symmetry, giving rise to novel physical properties, but also posing significant experimental challenges in ambiguously identifying intrinsic multiferroicity. This review critically examines the current landscape of 2D multiferroic materials, with an emphasis on experimental methodologies rather than material taxonomy. We assess conventional probes for magnetism and ferroelectricity, identifying key limitations in single-technique approaches. We emphasise the importance of magneto-opto-electric cross-correlative measurements in establishing intrinsic multiferroic behaviour and outline future research directions for both fundamental studies and device applications.**

**Key points**

- **2D vdW single-phase multiferroics offer unprecedented opportunities for magnetism, ferroelectricity, and their coupling at the atomic scale, enabling external control and switching via electric fields, among other applications.**

- **Simultaneous breaking of time-reversal and spatial-inversion symmetry – a hallmark of single-phase multiferroics – poses fundamental challenges for experimental identification, often invalidating techniques that assume independent ferroic orders.**

- **Conventional characterisation methods that target only magnetic or ferroelectric orders can yield misleading signals when cross-coupling is present, leading to false positives.**

- **Cross-correlated magneto-opto-electric measurements are crucial for disentangling intertwined ferroic responses, revealing mutual control between magnetic and electric orders, and establishing intrinsic multiferroicity in 2D materials.**

**Box 1 | Symmetry breaking in 2D vdW multiferroics**

**Spatial-inversion symmetry breaking due to ferroelectric polarizations**

Ferroelectricity in 2D materials fundamentally breaks spatial-inversion symmetry normally arises through two mechanisms: the ionic-displacement within monolayers (e.g., $CuInP_2S_6$) and interlayer charge redistribution caused by lateral sliding in multilayer stacks (e.g., bilayer *h*-BN). Artificially stacked layers in twisted moiré systems with small twist angles (e.g., twisted graphene/*h*-BN) can form moiré patterns with alternating polarization domains, where the local stacking structure directly controls spatial-inversion symmetry breaking. Collectively, these atomic-scale symmetry-breaking mechanisms enable unprecedented control over ferroelectric functionalities in low-dimensional systems.

**Time-reversal symmetry breaking due to magnetic orders**

Ferromagnetic order breaks time-reversal symmetry through a net magnetic moment ($\boldsymbol{M} \neq 0$), which flips all spins ($\boldsymbol{S_i} \rightarrow -\boldsymbol{S_i}$) and reverses $\boldsymbol{M} \rightarrow -\boldsymbol{M}$ under *T*, resulting in a distinct state separated by an energy barrier. In 2D ferromagnets, the magnetic exchange interactions always result in intralayer/interlayer parallel spin alignments. In antiferromagnets, *T*-symmetry is broken by the ordered spin sublattices, characterized by the *T*-odd Néel vector $\mathbf{M}_1-\mathbf{M}_2$. Considering a bipartite antiferromagnet with antiparallel spins (A: ↑, B: ↓), time reversal flips all spins (A: ↓, B: ↑), equivalent to swapping sublattices A and B. Since actual materials always exhibit broken sublattice symmetry via anisotropy or Dzyaloshinskii-Moriya interactions, the time-reversed state is inequivalent to the original and thus time-reversal symmetry is broken despite zero net magnetic moment.

**Simultaneous breaking of time-reversal and spatial-inversion symmetry**

In layered antiferromagnets such as bilayer $CrI_3$, both spatial-inversion operation and time-reversal operation in the antiferromagnetic bilayer convert one layered antiferromagnetic state to the other, but inequivalent to itself, and thus time-reversal and spatial-inversion symmetry are broken simultaneously. Specifically, in each ferromagnetic monolayer, spins are aligned in parallel, while the adjacent layers couple antiferromagnetically with opposing spin orientations. This magnetic configuration is not invariant under time reversal, which would flip all spins, transforming into a different magnetic state. Simultaneously, an spatial inversion operation would also reverse the spin directions due to the antiferromagnetic interlayer coupling, resulting in a distinct magnetic configuration rather than the original one. This dual symmetry breaking originates purely from the magnetic order, independent of any structural non-centrosymmetry, since the atomic lattice of bilayer $CrI_3$ itself is centrosymmetric.

In helimagnetic systems like $NiI_2$, spins at neighboring sites within a monolayer arrange themselves in a helical pattern and break time-reversal and spatial-inversion symmetry through chiral spin textures. Both time reversal and spatial inversion operation can flip all spins ($\boldsymbol{S}_i \rightarrow -\boldsymbol{S}_i$), transforming a right-handed spin spiral into a left-handed one ($\boldsymbol{\kappa} = \boldsymbol{S}_i \times \boldsymbol{S}_j$). The inverse DMI effect generates ferroelectric polarization through $\boldsymbol{P}_{ij} = A\boldsymbol{e}_{ij} \times (\boldsymbol{S}_i \times \boldsymbol{S}_j)$, resulting in robust intrinsic type-II multiferroic orders.

Multiferroic materials, characterized by the coexistence and coupling of multiple ferroic orders (e.g., ferroelectricity, ferromagnetism, and ferroelasticity), have garnered substantial research attention due to their unique potential in next-generation device applications[1-3]. The rich physical coupling between these ferroic orders—manifested through magnetoelectric, piezoelectric, and magnetoelastic interactions—gives rise to unprecedented functionalities, including electric-field-controlled magnetism and strain-mediated polarization switching. The hallmark of multiferroics lies in their magnetoelectric coupling, where magnetic properties can be controlled by electric fields and vice versa, offering a pathway to overcome the energy efficiency limitations inherent in conventional current-driven magnetic switching technologies[4,5]. Generally, the realization of strong magnetoelectric coupling in single-phase materials remains challenging due to competing electronic requirements (e.g., the $d^0$ rule for ferroelectricity versus partially filled *d*-orbitals for magnetism) and symmetry constraints[6]. While pioneering works on three-dimensional (3D) multiferroics, such as $BiFeO_3$ (type-I multiferroic) and rare-earth manganites $TbMnO_3$ (type-II multiferroic), have demonstrated remarkable magnetoelectric effects[7-9], their practical applications have been hindered by limited magnetoelectric coupling strength and device integration challenges.

However, the recent advent of van der Waals (vdW) magnets that show stable magnetism on atomically thin samples offers a completely new opportunity in two dimensions[10-17]. Unlike their bulk counterparts, two-dimensional (2D) vdW multiferroics represent a groundbreaking class of materials that simultaneously exhibit multiple ferroic orders within a single atomic layer, offering unique advantages such as enhanced magnetoelectric coupling, remarkable design flexibility and exceptional tunability. The reduced dimensionality in these materials significantly amplifies the coupling between charge, spin, and lattice degrees of freedom, enabling unprecedented control over ferroic orders via external stimuli such as electric fields, strain, and stacking engineering[18-20]. The field of 2D multiferroics has witnessed remarkable progress in recent years, with significant breakthroughs in both material discovery and property manipulation. Several innovation strategies have been developed to overcome the inherent incompatibility between ferroelectricity and magnetism in low-dimensional systems. For example, new 2D multiferroic systems have been realized by creating ferroelectric order in 2D magnets or establishing magnetic order in 2D ferroelectrics through interfacial engineering, chemical stoichiometry controlling and magnetic doping[21-30]. In this regard, it is important for us to begin with the recent discovery of intrinsic type-II multiferroic order in vdW $NiI_2$ at the limit of a few atomic layers. This material remarkably exhibits the coexistence of chiral spin textures, antiferroelectricity and ferroelectricity with strong magnetoelectric coupling[31-35], prompting extensive research interest in atomically thin flakes. In such type-II multiferroics, the direct coupling between the magnetic and ferroelectric order parameters is enabled by the presence of a spin configuration lacking inversion symmetry, resulting in a large and robust magnetoelectric response[8,36-41]. This makes 2D multiferroic materials exceptionally promising for advanced high-density and low-power spintronics and memory devices.

Despite significant advances in conventional characterization techniques for probing 2D magnetism (e.g., magneto-optical effects and anomalous Hall effect) and ferroelectricity (e.g., piezoresponse force microscopy (PFM), second harmonic generation (SHG) and linear dichroism (LD) spectroscopy), coupling between magnetic and ferroelectric orders results in the challenge of conventional methods for identifying multiferroicity in single-phase 2D multiferroic materials. Crucially, these single-technique approaches fail to directly capture magnetoelectric coupling—the defining feature of multiferroicity—or provide simultaneous insights into intertwined magnetic, ferroelectric, and electronic degrees of freedom. For example, the all-optical characterizations are not sufficient to prove 2D multiferroics in the presence of noncollinear and antiferromagnetic orders[32,33]. This inadequacy underscores the urgent need for developing innovative cross-correlative magneto-opto-electric joint-measurement technologies. Such an integrated approach should combine steady-state/time-resolved spectroscopy, electronic transport, and spin-resolved photocurrent measurements with unmatched spatiotemporal resolution (down to the nanometer and femtosecond scales) under multifield conditions (low temperatures, high magnetic/electric fields), thereby enabling the direct observation of multiferroic orders and magnetoelectric coupling in 2D systems.

This review provides a timely and in-depth survey of 2D multiferroics, with a focus on methodological advances in characterizing 2D multiferroicity. This review goes beyond merely listing characterization techniques that are only suitable for detecting isolated magnetic and ferroelectric orders. From the viewpoint of symmetry, this review clarifies the coupling relationships among magnetic order, ferroelectric order, and optical signals, and disentangles magnetic/ferroelectric signatures from complex optical signals. This review also establishes the criteria and critical evaluation for distinguishing intrinsic multiferroicity in single-phase 2D multiferroic materials. Specifically, this review is structured first to briefly introduce 2D vdW multiferroics and the symmetry breaking therein, which are experiencing rapid advances, and then, focus on critical evaluation of techniques and pitfalls in distinguishing intrinsic multiferroicity (magneto-optical effect (MOKE, MCD/MLD), SHG, circular photogalvanic effect (CPGE), bulk photovoltaic effect (BPVE)), and establish the criteria for identifying intrinsic multiferroicity in single-phase 2D multiferroic materials. Finally, we outline the pathway for developing next-generation characterization tools to advance the fundamental understanding and practical applications of 2D multiferroics. This review, with a particular focus on the physics of "distinguishing intrinsic multiferroicity", is complementary to other recent reviews on this subject, which catalogue and discuss the discovered 2D multiferroic materials and their emergent applications.

## 1. Optical and electrical signatures due to symmetry breaking

### 1.1 Optical signatures induced by time-reversal symmetry breaking

2D magnetic materials have opened new avenues for studying low-dimensional magnetism and developing novel spintronic devices[42]. To probe their magnetic properties, several advanced magneto-optical techniques have been established, offering non-contact, high-sensitivity measurements with sub-micrometer resolution. These methods rely on detecting magnetic-order-induced optical dichroism, where the measurable quantities are related to the complex optical conductivity ($\sigma=\sigma_{real}+i\sigma_{imag}$, with $\sigma_{real}$ and $\sigma_{imag}$ denoting the real and imaginary parts) of the material[43-45].

The magnetic circular dichroism (MCD) effect, magneto-optical Faraday rotation (MFR) and magnetic linear dichroism (MLD) are three typical magnetic dichroism effects (Fig. 1), where MCD and MFR are highly sensitive to magnetization ($M$) and are analogous to the anomalous Hall effect (AHE), whereas MLD depends on $M^2$ and is analogous to anisotropic magnetoresistance in electrical transport. The MCD and MFR effect corresponds to the amplitude difference (differential absorption) and phase difference of left-circularly and right-circularly polarized light (Fig. 1b), arising from the antisymmetric off-diagonal components of the optical conductivity tensor. One experimental scheme to carry out MFR and MCD experiments is to measure changes in the rotation angle and ellipticity of the polarization for the reflected or transmitted light[46], which are proportional to the real and imaginary parts of $\sigma_{xy}$, respectively, and the rotation angle can be determined from the intensity differences between co- and cross-configuration linearly polarized light. Reflective MCD (RMCD), reflective MFR (RMFR) and reflective MLD (RMLD) combine the principles of MCD, MFR and MLD with reflection geometry, enhancing sensitivity for thin, opaque samples like 2D vdW magnets. The RMCD approaches have been crucial for investigating magnetism in 2D systems[14,19,47-52], such as in $Fe_3GeTe_2$, $CrI_3$ and $CrSe_2$. In contrast to RMCD, RMLD probes absorption differences for linearly polarized light parallel and perpendicular to the magnetization direction. Since MLD depends on $M^2$, the RMLD signal is thus insensitive to the sign of $M$. Figure 1c shows the corresponding RMCD and RMLD maps of a $NiI_2$ flake while sweeping the magnetic field from +1.0 T to -1.0 T at 10 K, with odd-symmetric RMCD and even-symmetric RMLD behaviors[32,53,54]. RMLD is also equally responsive to both ferromagnetic and antiferromagnetic orders, enabling the detection of collinear antiferromagnetic materials[55,56], such as $FePS_3$ and $NiPS_3$.

The Magneto-optical Kerr effect (MOKE) is another powerful optical technique for investigating magnetic properties based on the magnetic circular birefringence. It measures changes in the polarization rotation angle for linearly polarized incident light (linearly polarized light can be considered as an equal superposition of RCP and LCP). When linearly polarized light is normally incident and reflects off the magnetized material, both the phase and amplitude difference between RCP and LCP leads to a rotation of the linear polarization with a Kerr angle $\theta_K$ (Fig. 1b), which is proportional to the difference in the imaginary optical conductivity $\sigma_{imag}$[43,44]. The polar MOKE measurement[57] is the first to demonstrate the confirmation of robust FM orders in 2D magnets, such as in monolayer $CrI_3$ and bilayer $Cr_2Ge_2Te_6$, with clear magnetic hysteresis and a large remanent magnetization at zero magnetic field[11,12]. MOKE and RMCD have been widely used to detect magnetic properties in 2D magnets (such as $Ta_3FeS_6$, $Cr_5Te_8$, and $Fe_5GeTe_2$)[58-61] due to its high sensitivity and capability to probe magnetization dynamics and domain structures. In essence, RMCD, RMFR, RMLD and MOKE are equivalent, which are simply four different manifestations of the same underlying magnetism (Fig. 1b). These optical techniques are ideal for detecting small and weak signals from as-exfoliated 2D flakes without involving any device fabrication process, which might perturb the intrinsic magnetism.

### 1.2 Optical signatures induced by spatial-inversion symmetry breaking

The inherent breaking of spatial-inversion symmetry in materials, including non-linear crystals and ferroelectrics, induces pronounced crystal anisotropy both along and perpendicular to the polar axis, resulting in distinct optical signatures. This anisotropy can be quantitatively characterized through linearly polarized light spectroscopy, typically yielding either linear dichroism (LD)—defined as the differential reflectance between two different polarizations—or birefringence ($\Delta n$), representing the refractive index contrast between two crystallographic axes[55,62], as shown in Fig. 2a-b. In ferroelectrics, the electronic transitions are polarization-dependent, leading to differential absorption/reflection of light polarized parallel or perpendicular to the ferroelectric axis. Polarization-resolved transmission spectroscopy is commonly employed to derive LD spectra, while the optical constants (refractive index $n$ and extinction coefficient $k$) can be extracted through the transfer matrix method[55,63]. And the birefringence is experimentally verified using polarization-resolved optical microscopy with crossed polarizers. In birefringent materials, a relative phase shift is introduced between orthogonal polarization components, resulting in elliptical polarization that partially transmits through the analyzer system. In contrast, the polarized light in an isotropic material experiences no phase retardation and is completely extinguished by the analyzer. For example, various 2D ferroelectric materials exhibit strong LD effect, and $NbOCl_2$ flakes display near-unity LD contrast (up to 99%) in the ultraviolet regime[64], as shown in Fig. 2c. This extreme anisotropy arises from its highly dispersive conduction band along one crystallographic direction and flat bands along the orthogonal axis. The refractive index $n$ along the two in-plane axes also shows a remarkable contrast, demonstrating a large in-plane birefringence ($\Delta n$ = 0.26–0.46) across a broad transparency window from visible to near-infrared (Fig. 2d), attributed to Peierls distortion induced Nb dimerization. This structural distortion localizes valence electrons and enhances polarizability along a specific axis, underpinning its remarkable optical anisotropy. However, in multiferroic systems, both magnetic orders ($M$) and ferroelectric polarizations ($P$) can lead to intensity differences of polarized light between parallel and perpendicular configurations. The LD/BF signals arising from ferroelectric structural asymmetry and the MLD/MFR originating from magnetization always exhibit similar experimental manifestations in polarization-dependent absorption or reflection, making it difficult to distinguish whether the observed anisotropy originates from ferroelectric or magnetic orders without further symmetry-based analysis or external field manipulation.

Second-harmonic generation (SHG) is a second-order non-linear optical process wherein two incident photons of frequency $\omega$ coherently combine within a non-centrosymmetric medium to generate a single photon at double the frequency ($2\omega$), as shown in Fig. 2e. This phenomenon is symmetry-forbidden in centrosymmetric crystals, making the detection of SHG a direct probe of broken inversion symmetry and, consequently, a highly sensitive indicator of ferroelectric orders. The intensity of SHG emission scales quadratically with the material's second-order non-linear susceptibility tensor $\chi^2$. Crucially, $\chi^2$ is intrinsically linked to spontaneous polarization in ferroelectrics, as both of them arise from non-centrosymmetric charge distributions. Such SHG technique is widely used for detecting symmetry breaking in both 2D in-plane and out-of-plane ferroelectric materials in recent studies[65-69]. For instance, 2D in-plane ferroelectric $NbOI_2$ nanosheets exhibit record SHG conversion efficiencies (>0.2%), attributed to robust in-plane ferroelectricity that maximizes $\chi^2$ components (Fig. 2f-g). Note that LD or BF merely requires anisotropic optical properties in materials, without any intrinsic dependence on ferroelectric order, as observed in many 2D anisotropic materials[63,70-73]. Likewise, SHG fundamentally relies on non-centrosymmetric crystal structures; however, the absence of inversion symmetry does not universally imply ferroelectric behavior, as exemplified by non-ferroelectric monolayer transition metal dichalcogenides (TMDCs). Moreover, pure magnetic orders can induce strong SHG or LD signals, which have been observed in many magnetic materials[55,74-76]. Consequently, both LD and SHG measurements should be considered only as secondary indicators when characterizing ferroelectric materials.

### 1.3 Correlations of optical signatures and symmetry breaking

Although all-optical characterizations provide powerful probes of symmetry breaking, they sometimes cannot constitute definitive proof of ferroelectricity in single-layer multiferroics where magnetic and ferroelectric orders coexist. In non-centrosymmetric systems, SHG is permitted through electric-dipole transitions and exhibits time-reversal symmetry, classified as i-type SHG. Conversely, even in centrosymmetric crystals, SHG can occur when magnetic ordering simultaneously breaks both spatial inversion (*P*) and time-reversal (*T*) symmetries[77-80]. This magnetically-induced SHG process, designated as c-type, is characterized by time-non-invariant and nonreciprocal behavior. For instance, centrosymmetric antiferromagnetic materials like bilayer $CrI_3$ exhibit strong SHG signals purely from magnetic-order-induced symmetry breaking, independent of any ferroelectric polarization[81]. Besides collinear antiferromagnetic systems[81,82], noncollinear spin arrangements can also generate vector spin chirality ($\kappa = \mathbf{S}_1 \times \mathbf{S}_2$) that fundamentally alters symmetry properties. This chiral magnetic order can simultaneously break both spatial inversion (*P*) and *PT* symmetries, giving rise to an additional i-type SHG response of purely magnetic origin (chiral-SHG), as demonstrated in the vdW CrSBr magnet[83].

On the other hand, while LD and BF typically probe ferroelectric anisotropy, it can also emerge from magnetically induced lattice distortions in antiferromagnetic systems. A striking example is the vdW antiferromagnetic insulator $FePS_3$, where zigzag spin ordering below Néel temperature generates substantial in-plane anisotropy through strong spin-charge coupling. It is worth noting that $FePS_3$ is the first vdW magnet shown to have stable magnetism on a monolayer[10]. This magneto-optical effect produces exceptional LD signals and birefringence comparable to those found in ferroelectric materials[55]. These observations reveal a fundamental limitation in the all-optical characterization of multiferroics: neither SHG nor LD signals can unambiguously confirm ferroelectricity, particularly when antiferromagnetic and noncolinear orders coexist. Definitive verification of ferroelectricity therefore requires field-dependent measurements–applying electric fields to test polarization switching and magnetic fields to decouple magneto-optical contributions to distinguish between structural and magnetic origins of the optical anisotropy. For example, the observed SHG and LD in $NiI_2$ could, in principle, stem from helical magnetic order alone in principle, making the claim of ferroelectricity harder to accept without additional electric-field-dependent validation[31,32]. The ferroelectricity in single-phase multiferroics must be validated by the electric/magnetic field control of magnetism and ferroelectric polarization[33,84].

Table 1 summarizes the high diversity of optical and ferroelectric parameters in existing two-dimensional nonlinear and multiferroic materials. Optical responses such as strong LD and large second-order nonlinear susceptibility ($\chi^{(2)}$) are widely observed in many non-ferroelectric two-dimensional nonlinear materials (e.g., black phosphorus, $ReS_2$) and magnetically ordered systems (e.g., $FePS_3$). Meanwhile, these materials can form heterostructures with 2D materials, for instance, emerging wurtzite nitride ferroelectrics (e.g., $Al_xSc_{1-x}N$ exhibit $P_r$ > 100 μC/cm² with tunable $E_c$ (2–8 MV/cm)) can be integrated with 2D materials to create high-performance heterojunctions[85,86], which may exhibit comparable optical behavior. This indicates that among numerous 2D nonlinear/ferroelectric/multiferroic materials and their heterostructures, relying solely on a single optical characterization technique (such as LD and SHG cannot clearly distinguish whether the signal originates from symmetry breaking in the ferroelectric structure, lattice anisotropy induced by magnetic order, or the intrinsic nonlinear optical effects of the material.

## 1.4 Electrical signatures induced by symmetry breaking

Beyond optical characterization, electrical transport measurements have emerged as powerful tools for investigating magnetic states in two-dimensional materials. The anomalous Hall effect (AHE) represents a particularly sensitive probe of broken time-reversal symmetry in magnetic systems, arising from extrinsic (e.g., skew scattering and side-jump effects caused by impurity scattering) and intrinsic (governed by Berry curvature in momentum space, reflecting the topological nature of electronic bands) mechanisms[87]. Distinct from the ordinary Hall effect that requires external magnetic fields, the AHE manifests spontaneously as a transverse voltage in the absence of external magnetic fields through either intrinsic magnetization or noncollinear spin textures (Fig. 3a). The Hall resistivity $\rho_{xy} = R_0B + R_sM,$ where $R_0$ is the ordinary Hall coefficient driven by Lorentz force, $R_s$ is the anomalous Hall coefficient and $M$ denotes the out-of-plane magnetization. This unique sensitivity to magnetization makes AHE particularly valuable for studying ultrathin magnetic films, where traditional magnetization measurements are difficult to probe. In metallic ferromagnets like $Fe_3GeTe_2$, AHE measurements have first conclusively demonstrated the ferromagnetic order down to the monolayer limit (Fig. 3b)[13]. AHE is then be used to probe magnetic orders in $Fe_4GeTe_2$ and $Fe_5GeTe_2$ as well as in many 2D magnets, establishing it as a definitive probe of 2D ferromagnetism[15,60,88-90]. This capability is particularly powerful for investigating magnetic anisotropy, as demonstrated by the gate-induced spin-flop transitions from out-of-plane to in-plane in $Fe_5GeTe_2$ through AHE measurements[60]. More remarkably, recent advances have extended AHE studies to two-dimensional antiferromagnetic systems[91], opening new avenues to explore topological electronic bands arising from non-trivial Berry curvature and complex spin textures at 2D limit.

Ferroelectric polarization can be characterized by measuring charge-voltage (*Q-V*) hysteresis loops or polarization-electric field (*P-E*) loops under alternating electric fields. This methodology probes the material's ability to exhibit switchable dipole moments, serving as a fundamental criterion for identifying ferroelectricity. Under an applied AC field, the displacement current ($I=dQ/dt$) and capacitance ($C=\varepsilon A/d$) is measured and integrated to obtain the polarization charge ($Q=2P_rA$ for an ideal ferroelectric, where $P_r$ denotes the remanent polarization and $A$ represents the electrode area). For out-of-plane polarized ferroelectrics, the material is typically fabricated into a standard parallel-plate capacitor structure sandwiched by two electrodes (Fig. 3c). This configuration is particularly important for effectively eliminating parasitic effects arising from the nonuniform electric field or notably contact electrostatic artifacts generated by those tip-based approach (such as PFM)[92]. This approach has been successfully employed to demonstrate robust out-of-plane ferroelectricity in $NiI_2$ trilayer (Fig. 3d)[33] and giant negative piezoelectricity in layered ferroelectric materials[93] such as $CuInP_2S_6$. A horizontal geometry of electrode/ferroelectric/electrode is typically used for in-plane polarization ferroelectrics. For example, the ferroelectricity of $NbOI_2$ has been confirmed through $P-E$ hysteresis measurements conducted along its in-plane polar axis[65]. The obtained hysteresis loop exhibits well-defined switching behavior (Fig. 3d) with a coercive field of 8.5 kV cm$^{-1}$ and saturation fields of ±17.5 kV cm$^{-1}$, providing clear evidence of robust ferroelectric ordering in $NbOI_2$. Interestingly, in contrast to optical methods which may yield analogous signals stemming from either ferroelectric or magnetic phases, such P-E hysteresis measurements provide definitive evidence for ferroelectric orders in multiferroic materials since they directly demonstrate the reversible switching of electric dipoles under electric fields.

From this perspective, electrical characterization serves as a highly reliable and conclusive means of identifying ferroelectric order in multiferroic systems.

## 2. Magneto-opto-electric cross-probes of 2D vdW multiferroics

### 2.1 Steady-state magneto-opto-electric joint-measurement methods

Despite significant advances in conventional techniques for characterizing 2D magnetism (e.g., magneto-optical effects or anomalous Hall effect) and ferroelectricity (e.g., SHG, LD spectroscopy), these methods face challenges to identify multiferroicity in single-phase 2D multiferroic materials due to their inability to disentangled from magnetic and ferroelectric features complex signals, moreover failing to directly probe magnetoelectric coupling—the hallmark of multiferroic systems. For example, LD domains in $NiI_2$ flakes were observed to synchronize with magnetic domain switching under external magnetic fields (Fig. 1c)[32], raising questions about the exclusive contribution of ferroelectricity in this phenomenon. To unambiguously confirm the intricate multiferroic behavior, advanced experimental techniques capable of simultaneously probing magnetism and ferroelectricity with applying magnetic and electric fields are essential, highlighting the critical need for developing innovative cross-correlative magneto-opto-electric measurement technologies.

Recent studies have demonstrated that steady-state magneto-opto-electric joint-measurement methods are powerful tools for characterizing multiferroics (Fig. 4a), particularly through the development of advanced micro-nano scanning imaging systems that enable precise identification of low-dimensional multiferroic materials. Here, we highlight two key approaches: (a) electrically controlled RMCD and MOKE, and (b) magnetically controlled capacitance measurements. The integration of electric field control with RMCD/MOKE allows for direct observation of magnetoelectric coupling in 2D systems. The ferroelectric polarization can be nonvolatility switched by applying a voltage, which subsequently modulates the magnetic orders—such as exchange bias, magnetic anisotropy, coercive field, or spin-flip transitions—in multiferroics. These electrically induced magnetic changes are directly manifested as measurable alterations in the RMCD or MOKE hysteresis loops, such as horizontal shifts, changes in loop asymmetry, variations in coercivity, and the emergence of intermediate magnetic states, thereby enabling real-time characterization of magnetoelectric coupling phenomena.

A prominent example is demonstrated in the multiferroic $CrI_3$/α-$In_2Se_3$ heterostructure[94]. A 10 V gate voltage switches the ferroelectric polarization of α-$In_2Se_3$, which nonreciprocally modulates the interlayer magnetic coupling in $CrI_3$, resulting in a significant reduction of the spin-flip field by up to 82%, confirming the role of ferroelectric polarization in modulating magnetic order (Fig. 4b). And all-electric field-controlled magnetic switching via magnetoelectric coupling is demonstrated in a trilayer-$CrI_3$/$HfO_2$ multiferroic system (Fig. 4c)[95,96]. The electric field control induces a colossal exchange bias modulation from –0.15 to +0.33 T, which is directly related to an asymmetric shift in the RMCD hysteresis loop. Further RMCD measurements indicate that this effect originates from electric-field-induced charge doping, which switches the bottom $CrI_3$ bilayer from antiferromagnetic to ferromagnetic orders and enhances the direct exchange coupling[97]. More generally, electric-field-controlled RMCD/MOKE measurements have been employed to demonstrate the electrical manipulation of magnetism in 2D magnets[19,50,98] such as $CrI_3$ and $Cr_2Ge_2Te_6$. This approach nevertheless serves as a powerful tool for investigating multiferroic phenomena.

Since the magnetic field can influence the ferroelectric properties in multiferroics, integrating *P-E* measurements with magnetic field control provides valuable insights into magnetoelectric coupling and can be a critical tool for studying 2D multiferroics. For example, magnetically controlled *P-E* measurements directly revealed the coexistence of ferroelectricity and antiferromagnetism in multiferroic $NiI_2$[32,33]. In such experiments, the sample is typically fabricated into a vdW heterostructure (e.g., graphene/hBN/$NiI_2$/graphene/hBN) and subjected to alternating electric fields while measuring polarization and current responses (Fig. 4a, enlarged part). The hBN encapsulation layer acts as an insulating barrier to suppress leakage currents while creating a linear/non-linear tandem capacitance structure, significantly enhancing the area-to-distance ratio in capacitance planes significantly (Fig. 4d), thereby enabling precise identification of the ultrathin $NiI_2$ layers[53]. In trilayer $NiI_2$, *P-E* and *I-E* hysteresis loops revealed the unprecedented coexistence of ferroelectric and antiferroelectric phases, revealing the magnetic control behaviors of the switching dynamics of the ferroelectric domains. Notably, an anisotropic modulation behavior of ferroelectric polarization by magnetic field has been observed (Fig. 4e-f), with around 7% suppression of the remanent polarization ($P_r$). Such magnetic control ferroelectric polarization reveals strong magnetoelectric coupling in $NiI_2$, where the noncollinear spiral spin configuration induces out-of-plane electrical polarization and magnetic fields mediate polarization reorientation through tilting of the spin rotation plane. On the other hand, in single-phase multi-ferroic such as $CuCrP_2S_6$, the magnetic field control of SHG have been used to probe magnetic field-induced electric polarization that reflects the magnetoelectric effect[99].

Steady-state magneto-opto-electric measurements, including electrically controlled RMCD/MOKE and magnetically controlled capacitance measurements, can provide direct evidence of magnetoelectric coupling by simultaneously probing the mutual control of electric and magnetic orders, thereby overcoming the limitations of conventional all-optical and uncoupled characterizations. Such comprehensive measurements not only resolve fundamental ambiguities in multiferroic behavior but also pave the way for unlocking the full potential of 2D multiferroics in next-generation spintronic and magnetoelectric devices.

### 2.2 Time-resolved cross-probing approaches

Beyond the steady-state magneto-opto-electric joint measurement approaches, time-resolved cross-probing techniques with femtosecond temporal resolution offer the capability to directly observe real-time interactions between electric polarization ($\Delta P$) and magnetization ($\Delta M$), providing a powerful tool for revealing the dynamic magnetoelectric coupling mechanism of 2D multiferroic materials. Theoretically, the magnetoelectric properties are governed by dynamical magnetoelectric coupling at finite frequencies, which can be described by the first-order magnetoelectric response[36,100] $\Delta P_i(\omega) = \alpha_{ij}(\omega)\Delta H_j(\omega)$, where $\alpha_{ij}(\omega)$ represents the complex dynamical magnetoelectric susceptibility tensor that quantifies the cross-coupling between modulated electric polarization $\Delta P_i(\omega)$ and the applied magnetizing field $\Delta H_j(\omega)$ in multiferroic systems. As a pioneering approach, time-resolved cross-probing techniques—particularly the combined implementation of time-resolved SHG (tr-SHG) and Kerr rotation (tr-Kerr) microscopy[36]—have enabled the first direct observation of coupled $\Delta P$ and $\Delta M$ oscillations, as well as the dynamical magnetoelectric coupling mediated by electromagnons (e.g., $EM_o$ and $EM_e$) in multiferroic $NiI_2$. A pump laser pulse serving as a weak perturbation to the ground state coherently excites an electrically active magnon mode, while a weak off-resonant probe pulse[101,102] monitors the resulting real-time modulations of dipolar and spin-spiral orders in real time SHG and Kerr channels. Remarkably, a robust π/2 phase shift was observed between the oscillations in these two measurement channels, confirming that the magnetoelectric effect originates from chiral symmetry breaking and the fundamental nature of the chiral magnetoelectric coupling in this system. The magnetoelectric coupling constant can thus be extracted. The obtained value ($Im(\alpha_{xx}) = 11 \times 10^3$ ps $m^{-1}$, far exceeding that of conventional helimagnets[103] like CuO (Table 1), providing direct evidence of enhanced magnetoelectric effects at 2D limit. These findings not only validate the theoretically predicted inverse Dzyaloshinskii-Moriya (IDM) mechanism but also highlight the synergistic effects of spin-orbit coupling and *d-p* hybridization as the physical origin of strong magnetoelectric coupling in $NiI_2$.

### 2.3 Opto-spintronic measurement methods

The bulk photovoltaic effect (BPVE) in non-centrosymmetric materials, which enables the direct conversion of light into DC current, has emerged as a prominent research focus in nonlinear optics[104-110]. The BPVE generates photocurrent through two primary mechanisms[111-113], the shift current arises from real-space displacement of electrons during inter-band transitions characterized by a shift vector associated with the Pancharatnam-Berry phase[104,112], and the injection current results from photoinduced changes in electron group velocity between energy bands. In nonmagnetic systems, these distinct photocurrent contributions can be differentiated by their characteristic polarization dependencies[112,113]. The linear shift photocurrent emerges when inversion symmetry breaking lifts the antisymmetric nature of the shift vector, enabling photoexcitation to produce a net real-space charge displacement and consequent DC current along ferroelectric polarization direction (top panel of Fig. 5a). Conversely, the circular injection photocurrent stems from the opposite Berry curvature distribution between $k$ and -$k$ points, which creates unequal optical transition probabilities under circularly polarized illumination, leading to an imbalanced $k$-space carrier distribution that generates uncompensated group velocities and a net current. Such switching of photocurrent directions with the helicity of circularly polarized light is also known as the circular photogalvanic effect (CPGE), as shown in top panel of Fig. 5b. The close connection between BPVE/CPGE and band geometric properties (quantum metric, Berry curvature) makes them powerful probes of material topology. Therefore, ferroelectrics that exhibit intrinsic charge polarization are expected to host BPVE and CPGE effects.

Deterministic electric control of spin helicity has been demonstrated in type-II multiferroics[84,103], where the handedness of the spin orders is strongly coupled with ferroelectric polarization. The left-handed (LH) and right-handed (RH) spin orders along with the ferroelectric polarization are mirror-symmetric, with polarization lying within the mirror plane. Note that the definitive evidences for spin-helicity must rely on the intensity difference ($\Delta I$ = ($I^{\mathrm{LH}}$- $I^{\mathrm{RH}}$) and the dichroism ($\Delta D$ = ($I^{\mathrm{LH}}$- $I^{\mathrm{RH}}$) / ($I^{LH}$+ $I^{\mathrm{RH}}$)) between LH and RH spin orders under the same ferroelectric polarization, and not merely reply on the reversal behavior of individual LH or RH screw components under opposite ferroelectric polarization, as shown in bottom panel of Fig. 5a-b. For multiferroic and ferroelectric materials, a significant pitfall requires particular attention: the BPVE arising from FE-induced electric field can be susceptible to being misled into misidentified CPGE due to their similar photo- responses under opposite electric field[114-116], particularly when measured along the polarization direction. Experimentally, BPVE generates a shift current under light excitation regardless of where the light is linearly or circularly polarized light. The ferroelectric polarization direction dictates the photocurrent direction, however the FE-switched photocurrent reversal in BPVE can spuriously resemble the CPGE response, particularly under circularly polarized light excitation, but $\Delta I$(BPVE) = 0 and $\Delta D$(BPVE) = 0 (bottom panel of Fig. 5a), which is distinct from the $\Delta I$(CPGE) >> 0 and $\Delta D$(CPGE) → 100% (bottom panel of Fig. 5b). Therefore, the determination of spin helicity cannot be based solely on the photocurrent reversal of LH/RH screw under positive/negative ferroelectric polarization, unless a pronounced $\Delta I$ and $\Delta D$ between LH and RH chiral spins under the same polarization $P$ is observed, which is the definitive evidence for identifying the helicity switching.

In 2D multiferroic materials, BPVE has been used as a sensitive local probe of ferroelectric orders. By applying a poling electric field to $NiI_2$ during cooling, a well-defined zero-bias photocurrent in the multiferroic phase has been observed with a direction aligned along the built-in electrical polarization (bottom panel of Fig. 5a). The magnitudes and polarity of this photocurrent reflect the net ferroelectric polarization. However, the BPVE usually cannot exhibit light polarization dependence, and the photocurrents remain almost constant as the excitation light transitions from LH to LP, and then to RH. Such an electrical switching of BPVE can only confirm the existence of ferroelectricity in $NiI_2$. Considering that the net ferroelectric polarization of multi-ferroics can be changed by an external magnetic field, magnetically controlled BPVE could serve as a powerful direct probe for investigating magnetoelectric coupling phenomena in multiferroic systems.

The CPGE effect was employed to probe the spin polarizations through ingenious cross-configuration photocurrent measurements analogous to the spin-valley Hall effect (Fig. 5b), which arises from spin-orbit coupling in its spin helix structure[117,118]. The comprehensive measured photocurrent can be expressed as[119] $J=C\sin(2\varphi)+L_1\sin(4\varphi)+L_2\cos(4\varphi)+D$, where $\varphi$ is the quarter-wave plate angle, $C$ denotes the spin photocurrent, $L_1$ and $L_2$ denote the linear photocurrent corresponding to the shift current and the thermal photocurrent and $D$ denotes a polarization-independent photocurrent[120,121]. Remarkably, when measured along the spin-antisymmetric direction perpendicular to the ferroelectric polarization, the $NiI_2$ system exhibited a pronounced helicity-dependent photocurrent difference between left (LCP) and right (RCP) circular polarizations. This observation provides evidence for the existence of spin textures. A crucial finding is that reversing the ferroelectric polarization via an electrical field induced an inversion of the spin photocurrent component, indicating the coupling between spin polarization and ferroelectric orders. In contrast, the photocurrent along ferroelectric polarization showed negligible CPGE, consistent with symmetric spin polarization along the $C_{2x}$ axis. Similarly, an organic–inorganic layered ferroelectric material with spin helix also demonstrated that the spin-polarized band structure is switchable by an ferroelectric field and a switchable CPGE[122].

The switching behaviors of both shift current and circular photocurrent are governed by the interplay among crystal symmetries, shift vector properties, and Berry curvature distributions. BPVE and CPGE responses in unknown material systems often involves complex mechanisms that require careful interpretation. Moreover, CPGE exhibits a pronounced excitation-wavelength dependent behavior, and a significant CPGE signal only emerges on resonant excitation at the band-edge (Fig. 5d-g), similar to the to the optical selection rule of valleytronics in monolayer TMDs[123]. The inter-/intra-valley spin relaxations drastically weaken the CPGE under off-resonance excitation. Recent theoretical and experimental advances have revealed richer BPVE phenomena in magnetic systems with different symmetry constraints: in magnetic materials where time-reversal symmetry (T) is broken but space-time inversion symmetry (PT) is preserved[124-126], photocurrent components including linear injection current and circular shift current emerge; while in systems lacking both T and PT symmetries[113,127], photocurrents can be generated by either linearly or circularly polarized light through additional symmetry-allowed mechanisms. These findings underscore the importance of comprehensive symmetry analysis when interpreting BPVE and CPGE in unknown material systems, particularly those involving complex magnetic orders.

In multiferroic and helimagnetic systems, photogenerated currents can effectively couple with inherent spin textures, providing a powerful approach for probing spin-related phenomena. The underlying mechanism involves light-induced excitation generating a charge photocurrent along a specific direction (e.g., the $xx$ direction). Due to strong spin–orbit coupling and non-trivial spin structures in such materials, this photocurrent might drive a transverse Hall voltage along the perpendicular direction (e.g., the $xy$ direction) through either the anomalous Hall effect or the spin Hall effect (Fig. 5c). This transverse signal directly manifests the coupling between spin and charge transport, with its amplitude and polarity being highly sensitive to both the spin configuration and the polarization state of incident light. For instance, under circularly polarized excitation, the mechanism becomes intrinsically linked to CPGE: the selective optical excitation of carriers with opposite chirality not only produces a longitudinal spin-polarized current but also generates a discernible transverse voltage by Hall effects, thereby enabling sensitive detection of spin textures and magnetoelectric coupling behavior.

## 3. Outlook

This review highlights methodological advances in characterizing 2D multiferroicity, where innovative cross-probing technologies are necessary for identifying intrinsic multiferroicity in single-phase multiferroics. Moving forward, enhancing the integration, sensitivity, and scalability of these cross-probing methodologies will be pivotal to advancing the field—from fundamental validation and discovery of novel multiferroic phenomena toward practical applications.

Beyond the current steady-state magneto-opto-electric measurements which focus on electrically controlled RMCD/MOKE and magnetically controlled capacitance measurements, future techniques should focus on integrating more complementary probes with enhanced spatial resolution. For example, integrating scanning probe techniques directly into magneto-optic and electrical measurement platforms enables real-space visualization of magnetoelectric coupling at nanoscale. This can be achieved through electric-field-controlled magnetic force microscopy (MFM) to image spin textures under bias, or magnetic-field-controlled piezoresponse force microscopy (PFM) to probe polarization switching in a magnetic field. Similarly, combining magnetically/electrically modulated optical scanning probes with in-situ transport, such as micro-scale OSHE/CPGE imaging measurements[128], can provide a direct way to detect interplay between ferroelectric and magnetic order within a single domain. As the field progresses towards room-temperature[129-137], wafer-scale ferroic/multiferroic films[138,139]—a prerequisite for applications—the capability of high-throughput operation these techniques will be important, which requires developing wide-field, high-sensitivity optical and scanning probe imaging (e.g., wide-field MOKE, SHG) combined with automated staging and machine-learning-based analysis to rapidly map the uniformity of ferroic orders across large areas. Concurrently, enhancing signal-to-noise ratios is critical for detecting weak multiferroic signals in emerging systems such as sliding ferromagnets[140-146]. The incorporation of ultrasensitive quantum sensors (notably nitrogen-vacancy (NV) centers in diamond[147-151]) into these vdW multiferroics can be a particularly promising research direction to meet this need, despite the technical challenge of such integration at cryogenic regime involved. On the other hand, the evolution of magnetoelectric coupling in multiferroic systems under dimensionality scaling can exhibit considerable complexity. For instance, in van der Waals systems such as $NiI_2$, multiple magnetoelectric phases, such as spiral multiferroic, interlayer collinear antiferromagnetic–antiferroelectric, and various metastable spin textures, are predicted to emerge due to dimensionality-driven reconstruction of magnetic exchange interactions, competing ground states, and substrate-mediated symmetry breaking[152]. Therefore, magneto-opto-electric measurement techniques will be essential for probing, distinguishing, and ultimately understanding these complex phase behaviors in low-dimensional multiferroics.

The time-resolved SHG/Kerr cross-probing technique, which combines micrometer-scale spatial resolution with fs-scale temporal precision, has enabled the in-situ observation of dynamical magnetoelectric processes within individual chiral domains. Notably, recent work demonstrates ultrafast polarization switching in sliding ferroelectric $WTe_2$ through sub-ns pulsing electrical excitations[153]. Looking forward, the integration of such ultrafast electrical excitation technique into magneto-optical pump-probe frameworks might unlock a new dimension of coupled magneto-photo-electric control. By synchronizing femtosecond optical pulses with programmable electrical gating, coherent multifield control over both electric and magnetic order parameters can be achieved on timescales that were previously inaccessible. This approach will unravel the fundamental speed limits of magnetoelectric switching and identify the primary bosonic mediators (phonons, magnons, electromagnons) of the magnetoelectric coupling. Pushing the spatial resolution of these techniques to the nanoscale via methods like time-resolved scanning near-field optical microscopy[154], will be essential for probing these dynamics within individual chiral domains or moiré unit cells, thereby clarifying how local structure governs ultrafast magnetoelectric response. Moreover, magnetically/electrically controlled BPVE and CPGE, may offer a unique way to disentangling the complex photocurrent contributions through symmetry analysis and multi-parameter experiments.

In conclusion, the continued development of cross-probing technologies will deepen our fundamental understanding of magnetoelectric coupling, validate emergent multiferroic properties and ultimately pave the way for practical applications. By combining these tools with scalable synthesis and precise engineering strategies, the field can overcome current limitations and unlock the full potential of 2D multiferroics for next-generation nanoelectronics, spintronics, and nonvolatile memory devices.

**Acknowledgments**

L.J.D., B.P., and H.T.Y would like to acknowledge the support by the National Science

Foundation of China (Nos. 52021001, 92365203, U24A6002), the National Key Research and Development Program of China (2024YFA1408104), and the Sichuan Provincial Outstanding Youth Science Foundation Project (2025NSFJQ0018), and the "Hundred Talents Program" Cultivation Project of UESTC (A1098531023601549), the Quantum Science and Technology-National Science and Technology Major Project (2021ZD0302502) and the Natural Science Foundation of Jiangsu Province (Nos. BK20253012, BK20243011, BK20233001). J.G.P. acknowledges the support by the Samsung Science & Technology Foundation (Grant No. SSTF-BA2101-05) and the Leading Researcher Program of the National Research Foundation of Korea (Grant No. RS-2020-NR049405).

**Author Contributions**

Bo Peng and Longjiang Deng conceived and designed the project, supervising all aspects. All authors co- wrote, discussed and commented on the manuscript.

**Competing interests:** The authors declare no competing interests.

**Table 1. Summary of optical and ferroic properties of various materials**

| Materials | LD | Δn | $\chi^{(2)}$ (pm/V) | $P$ (μC/cm$^2$) | $E_c$ (kV/cm) | $T_c$ (K) | EM | Refs. |
|---|---|---|---|---|---|---|---|---|
| BP | 15%~30% (VIS) | 0.245 (520 nm)<br>0.196 (642nm) | NA | NA | NA | NA | NA | 55,62 |
| $ReS_2$ | 5%~30% (VIS) | 0.037 (520 nm)<br>0.051 (642nm) | NA | NA | NA | NA | NA | 55,62 |
| $ReSe_2$ | 3%~10% (VIS) | 0.047 (520 nm)<br>0.082 (642 nm) | NA | NA | NA | NA | NA | 55,62 |
| $MoS_2$ | 42.4% (CBM, 2L) | 1.4 (1530 nm) | 311 (405nm, 1L) | NA | NA | NA | NA | 53,57,149,150 |
| 3R-$MoS_2$ | ~33% (860 nm)<br>~80% (430 nm) | 1.57 (760 nm)<br>0.97 (1520 nm) | 30 (600 nm) | 0.4 ($P_r$) | 360 | ~650 | NA | 57,151-153 |
| $NbOCl_2$ | 99% (UV) | 0.26-0.46 (VIS-NIR) | 400 (404 nm) | ~27 ($P_{sp}$) | ~4 | ~400 | NA | 55,57,154-156 |
| $NbOI_2$ | ~50% (470 nm) | 0.7 (1064 nm) | 190 (1050 nm) | ~18 ($P_{sp}$) | ~8.5<br>($t$ = 20 nm) | ~240 | NA | 56,154,156,157 |
| $CuInP_2S_6$ | ~80% (253.5 nm) | 1.24 (339.5 nm) | ~13 (1030 nm) | 4 ($P_r$) | 300 | 320 | NA | 153,158,159 |
| $CuCrP_2S_6$ | NA | 0.001 ($t$ = 61 nm) | NA | ~16 ($P_r$) | 250 | 32 ($M$)<br>333 ($P$) | NA | 22,25,153,160 |
| α-$In_2Se_3$ | ~20% (650 nm) | ~0.4 (568 nm)<br>~0.23 (482 nm) | 40 (800 nm) | 0.97 (OOP)<br>8.0 (IP) | 300 (OOP) | ~700 | NA | 57,153,161,162 |
| β-$In_2Se_3$ | ~20% (650 nm) | ~0.3 (~442 nm) | NA | 0.199 (DFT) | 0.27 eV/cell | ~470 | NA | 153,161,163 |
| $CrI_3$ | ~1% (RMCD) | ~5 mrad (MOKE, 633nm) | ~2000 (440nm) | NA | NA | ~45 | NA | 12,57,72,86-89 |
| $NiI_2$ | ~1.25% (532 nm) | ~15 mrad (532 nm) | ~43 (sum, 400 nm) | 0.1-0.3 ($P_{sat}$) | 500-750 | ~59 | 11×10$^3$ (ps/m) | 31,32,75,92,164 |
| $FePS_3$ | ~4% (633 nm) | 0.075 (500 nm) | ~0.13 (nm$^2$/V) | NA | NA | 117 | NA | 45-47 |
| $MgF_2$ | NA | 0.011-0.014 | NA | NA | NA | NA | NA | 165 |
| $CaCO_3$ | NA | 0.17 | NA | NA | NA | NA | NA | 166 |
| $AgGaS_2$ | NA | 0.044 (633 nm) | 23.6 (1054 nm) | NA | NA | NA | NA | 167 |
| $BaB_2O_4$ | ~20% (264 nm) | 0.12 (480 nm） | 2.2 (1064 nm) | NA | NA | NA | NA | 167 |
| $KH_2PO_4$ | ~80% (1064 nm) | ~0.04 (633 nm) | 0.39 (1064 nm) | NA | NA | NA | NA | 167 |
| $LiB_3O_5$ | NA | ~0.028 (532 nm) | 0.85 (1064 nm) | NA | NA | NA | NA | 167 |
| $ZnGeP_2$ | 32% (2.08 μm) | 0.075 (640 nm) | 75 (9.6 μm) | NA | NA | NA | NA | 167 |
| $LiNbO_3$ | ~30% (514 nm) | ~0.086 (588 nm) | 25.2 (1064 nm) | 71 | 20-50 | 1140 | NA | 167,168 |
| $HfO_2$ | NA | ~0.06 (633 nm) | 6 (800 nm) | 10-45 ($P_r$) | 1000-2000 | ~450 | NA | 169-171 |
| $Al_{1-x}Sc_xN$ | NA | ~0.02 (~1653 nm) | 42.5 (1550 nm) | ~70-130 ($P_r$) | 3000-5000 | ~1373 | NA | 77,78,172-175 |
| CuO | NA | NA | NA | 0.015 | 1 | ~230 | ~17 (ps/m) | 92,96 |
| $BiFeO_3$ | ~12% (XLD) | 0.275 (500 nm) | 38.4 (800 nm) | 100 ($P_{sp}$) | ~200 | ~1100 | ~3(V/cm·Oe) | 2,7,9,176,177 |
| $TbMnO_3$ | ~40% (633 nm) | ~0.5 (633 nm) | NA | 0.07-0.08 | 2 | ~28 | ~0.003@14T (μC/cm$^2$·T) | 8,178-180 |
| $DyMnO_3$ | NA | NA | NA | 0.3@2T | 20 | ~18 | ~0.09@7T (μC/cm$^2$·T) | 179 |

LD, linear dichroism; Δn, birefringence; $P_{sp}$, spontaneous polarization; $P_{sat}$, saturation polarization; $E_c$, Coercive fields; $T_c$, ferroelectric or ferromagnetic operating temperatures; VIS-NIR, visible to near-infrared; UVB-UVA, ultraviolet B to ultraviolet A; OOP, out-of-plane; IP, in-plane; XLD, X-ray linear dichroism; DFT, density functional theory calculation; NA, not applicable.

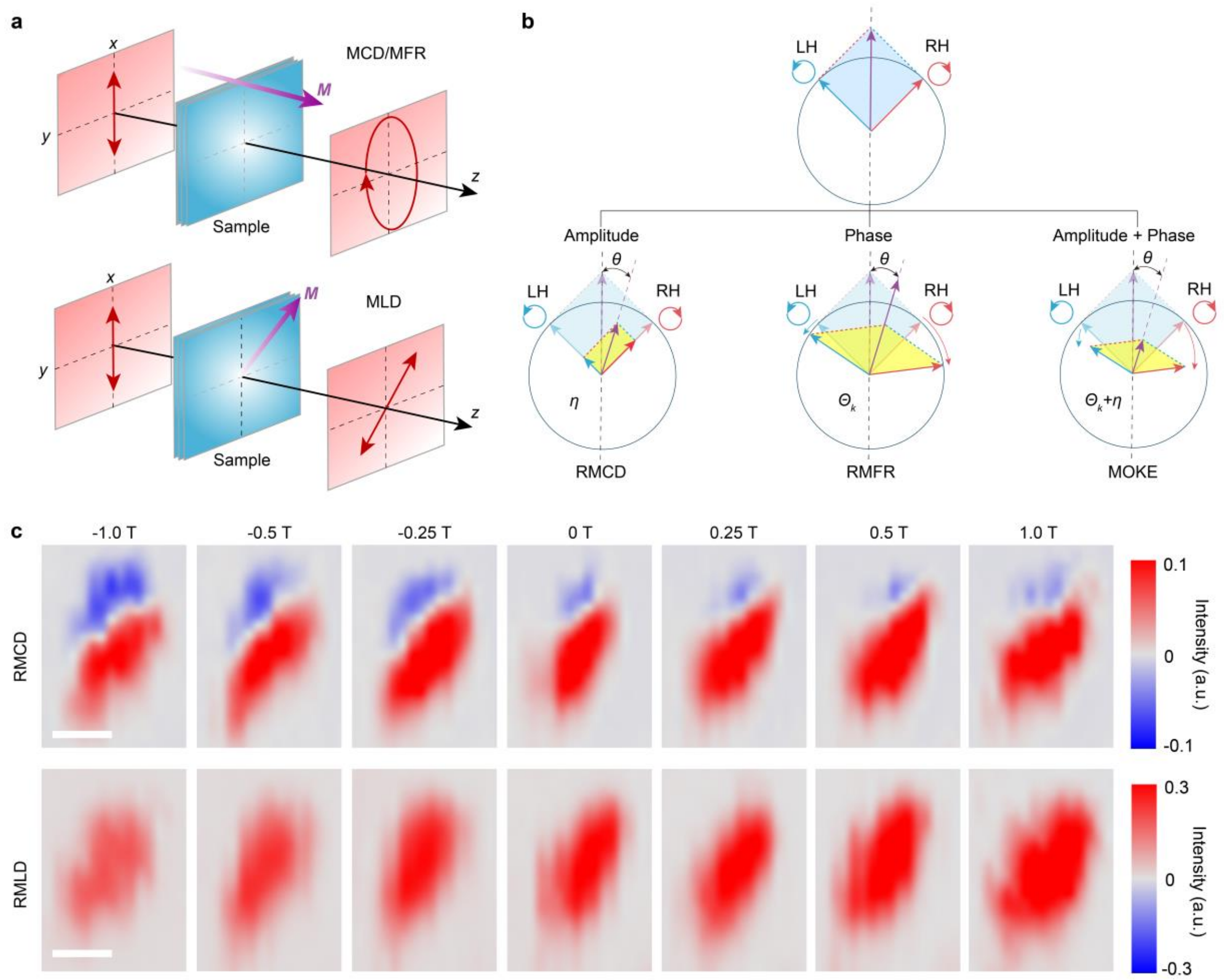

**Fig. 1** | Magnetic-order-induced optical signatures. **a** | Schematic of magnetic circular, Faraday rotation and linear dichroism (MCD/MFR/MLD). **b**| Reflective magnetic circular dichroism (RMCD), magneto-optic Faraday rotation (RMFR) and magneto-optic Kerr effect (MOKE). The linearly polarized light (solid violet arrows) can be decomposed into right-handed (light red arrows) and left-handed (blue arrows) circularly polarized components. RMCD and RMFR originate from the amplitude and phase difference between right-handed (RH) and left-handed (LH) polarized light, respectively, while MOKE results from the combined effects of both amplitude and phase differences. **c** | Corresponding two-dimensional RMCD and RMLD images in $NiI_2$ flakes at 10 K through our magneto-optical-electric joint-measurement scanning imaging system[53,54]. Scale bars, 1 μm.

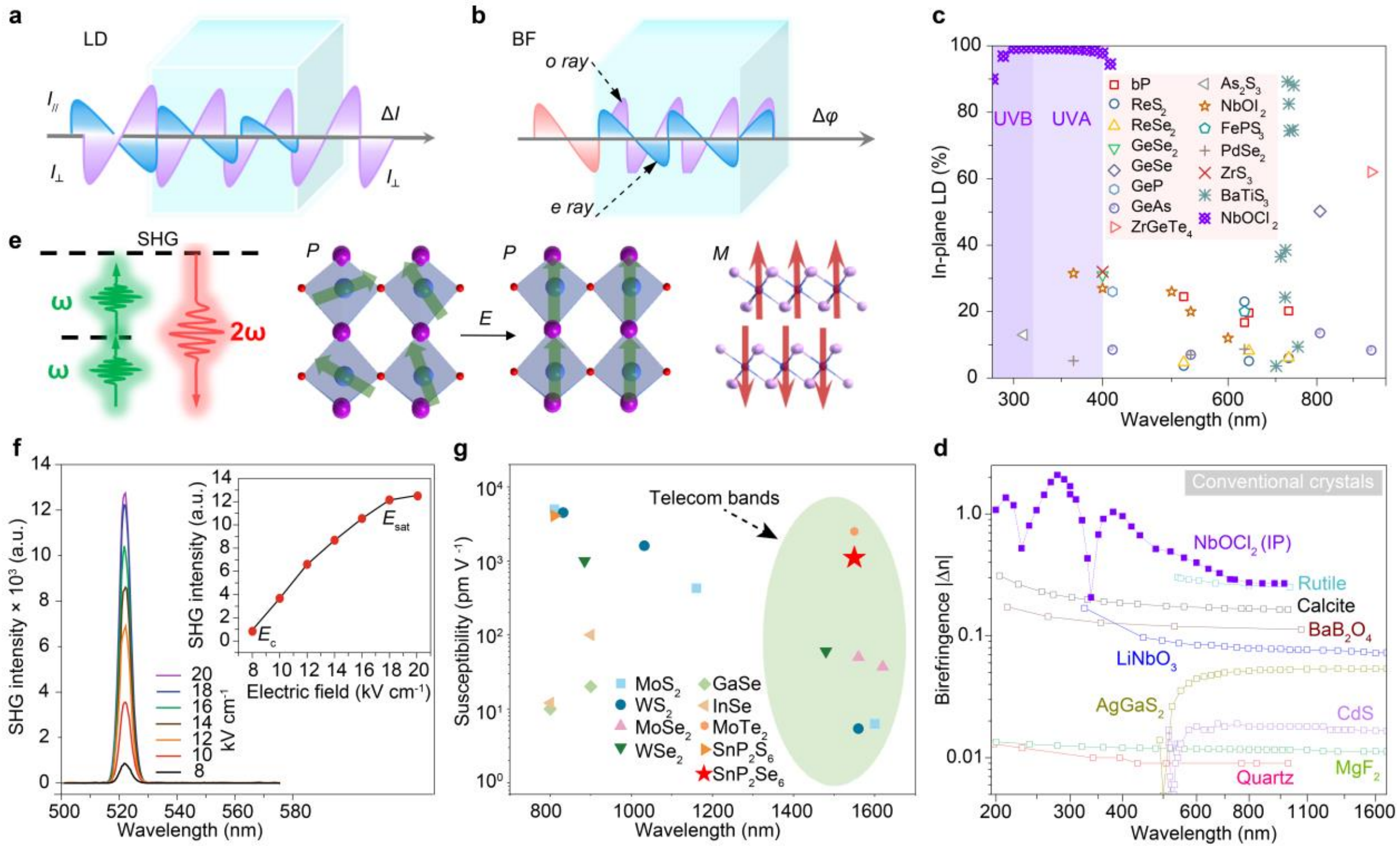

**Fig. 2** | Ferroelectric-polarisation-induced optical signatures. **a** | Linear dichroism (LD), **b** | Birefringence (BF), **c** | LD contrast in the ultraviolet and visible regime in diverse 2D materials, **d** | BF from visible to near-infrared range in a wide range of crystals. **e** | SHG due to the spatial-inverse symmetry breaking arising from ferroelectric polarizations and magnetic orders. **f** | Electric-filed-control of SHG features in $NbOI_2$ flakes. **g** | SHG efficiencies as a function of wavelength in diverse 2D materials. Panels c-d adapted from REF.[64], Springer Nature Limited. Panels f adapted from REF.[65], Springer Nature Limited. Panels g adapted from REF.[69], Springer Nature Limited.

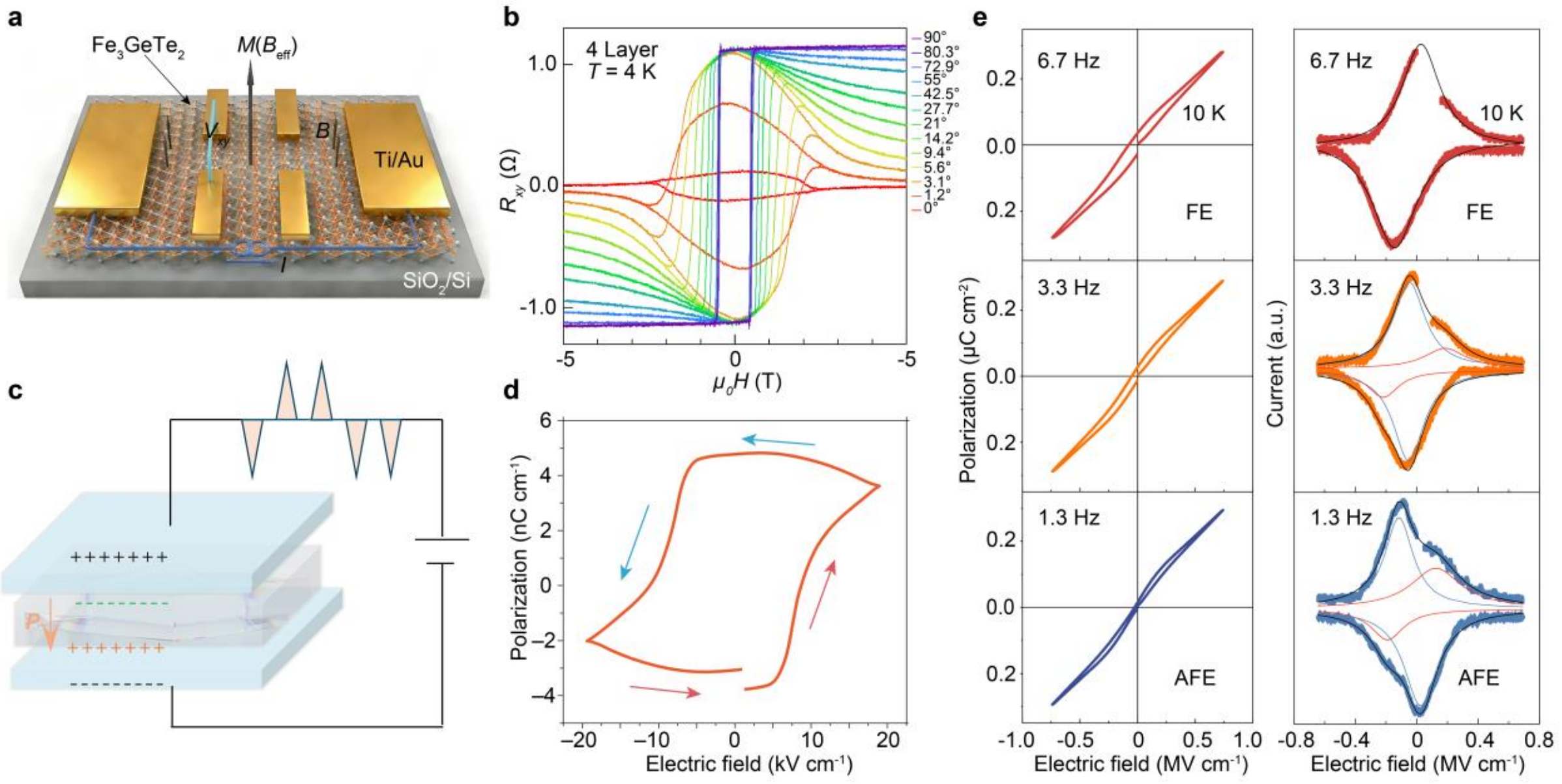

**Fig. 3** | Electrical measurement method. **a** | Schematic diagram of the device for detecting magnetic properties in 2D magnetism using anomalous Hall effect (AHE). **b** | Angle-dependent Hall resistance of a four-layer $Fe_3GeTe_2$ flake. **c** | Schematic picture of vertically aligned capacitance measurement technique for *P*-*E* hysteresis loops. **d** | P-*E* hysteresis loops of a $NbOI_2$ flakes in an in-plane configuration. **e** | P-*E* and corresponding *I*-*E* hysteresis loops at various frequencies in a trilayer $NiI_2$ vertical capacitance device. A double hysteresis loop and two pairs of current peaks (FE-AFE and AFE-FE switching peaks) were observed, indicating the coexistence of ferroelectric (FE) and antiferroelectric (AFE) polarization. Panels b adapted from REF.[13], Springer Nature Limited. Panels d adapted from REF.[65], Springer Nature Limited. Panels e adapted from REF.[33], Springer Nature Limited.

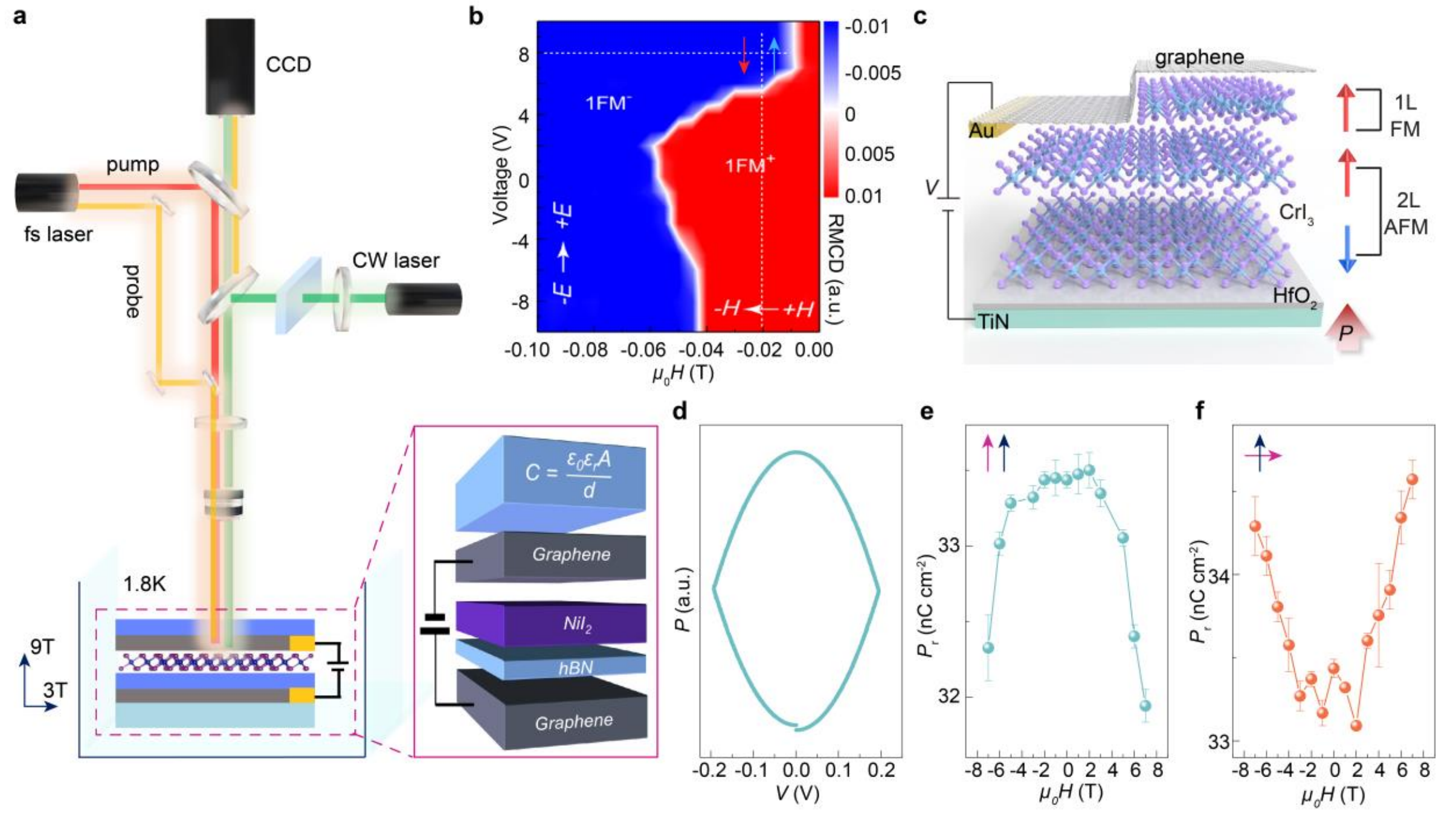

**Fig. 4** | Magneto-opto-electric cross-probes. **a** | Schematic of the steady-state/time-resolved magneto-opto-electronic cross-probe measurement system coupled with a *xz*-vector magnetic field (9-3 T) and a 1.8 K cryogenic system. The bottom right corner shows the schematic of a vertical trilayer $NiI_2$ device, sandwiched between graphene and h-BN. **b**,**c** | Electric-filed control of RMCD in trilayer $CrI_3/In_2Se_3$ and $CrI_3/HfO_2$ multiferroic heterostructure devices. **d** | P-V loop at 10 K from a $Gr/NiI_2/Gr$ device, showing serious leakage behaviours. **e**,**f** | Anisotropic magnetic control of ferroelectric polarization ($P_r$) in trilayer $NiI_2$ multiferroics as a function of in-plane magnetic field at H // *q*-vector and H ⊥ *q*-vector configurations, showing evident magnetoelectric coupling as a hallmark of 2D multiferroic. Panels b adapted from REF.[94], American Physical Society. Panels c adapted from REF.[95], Wiley-VCH. Panels d-f adapted from REF.[33], Springer Nature Limited.

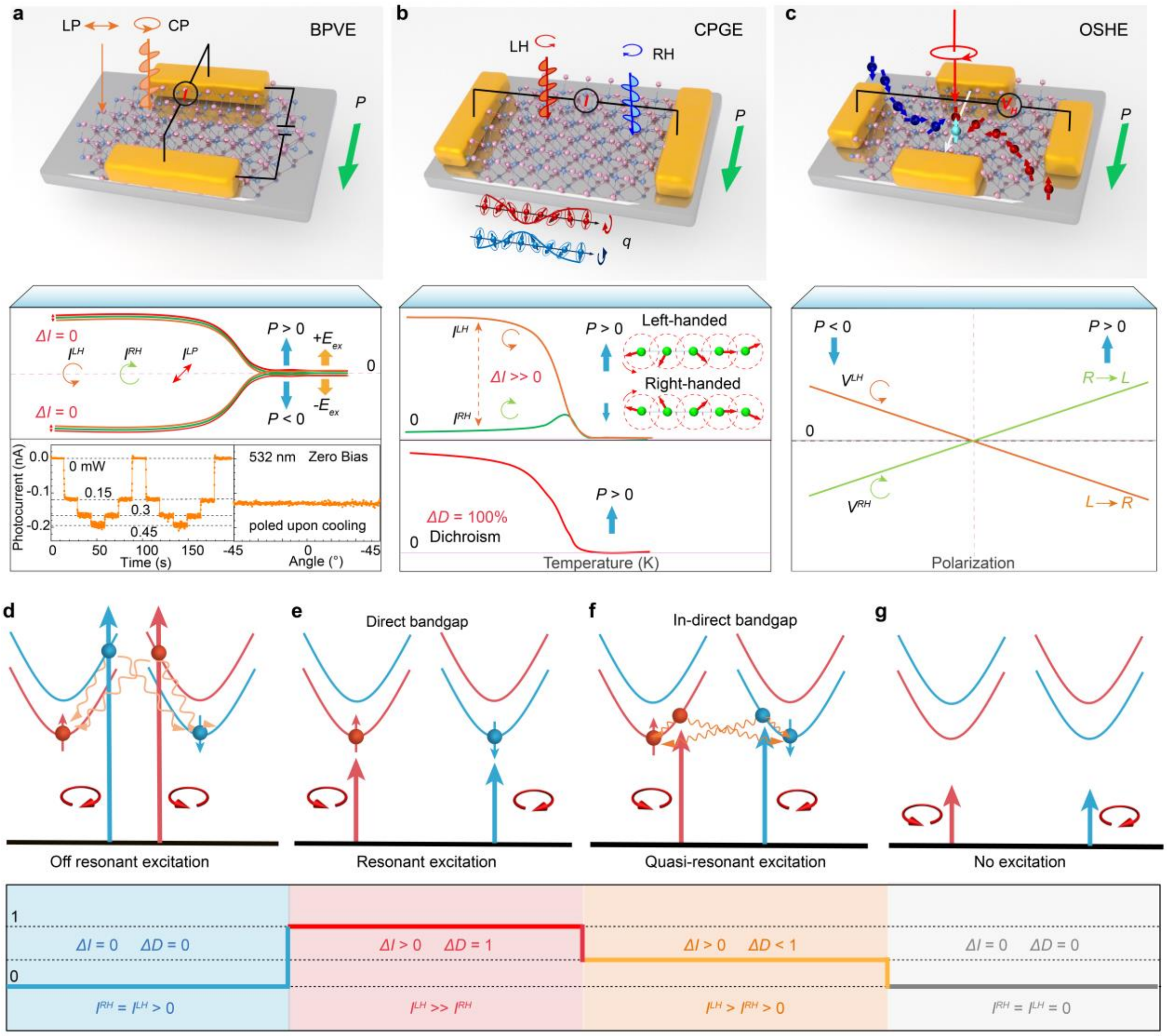

**Fig. 5** | Opto-spintronics. **a** | Schematic of bulk photovoltaic effect (BPVE). Top panel: BPVE schematic along with ferroelectric polarization. Bottom panel: BPVE schematic showing $\Delta I = 0$, the photocurrents do not show light-polarization dependence under the same ferroelectric polarization. The figure below shows the experimental BPVE results of bulk $NiI_2$ under 532 nm photoexcitation and zero-bias using our magneto-optical-electric joint-measurement scanning imaging system[53,54]. The photocurrents increase with the laser power increase from 0 to 0.45 mW, and vice versa, but the photocurrents show almost no change with light polarization. **b** | Schematic of circular photo-galvanic effect (CPGE). Top panel: of CPGE schematic with spiral spin orders and ferroelectric polarization. Bottom panel: CPGE schematic:

$\Delta I$ >> 0, showing strong circular polarization dependent photocurrent. **c** | Top panel: schematic of the optical spin hall effect principle (OSHE), with spin helicity ($M$), ferroelectric polarization ($P$) and light photons ($L$) together. Bottom panel: OSHE schematic showing the linear ferroelectric polarization ($P$) dependence of the Hall voltage ($V_H$) under the light polarization manipulation from left- to right-handed (orange solid line) and from right- to left-handed (green solid line). **d-g** | Wavelength dependence of CPGE: off-resonant excitation, on-resonant excitation with direct bandgap structures, quasi-resonant excitation with indirect bandgap structures, and no excitation.